# High-Throughput Computational Discovery of Inverted Resistive Switching in Two-Dimensional Materials

Sanchali Mitra[1], Arnab Kabiraj[2], Benjamin W. J. Chen[2], Han Zhang[3]*, Haiyu Meng[4], Shi-Jun Liang[4,5], C. S. Lau[1,6], Lei Shen[7], Lain-Jong Li[8], Kah-Wee Ang[9]*, Yee Sin Ang[1]*

[1] Science, Mathematics and Technology (SMT) Cluster, Singapore University of Technology and Design, Singapore 487372, Singapore

[2] Institute of Advanced Intelligence and Computing (IAIC), Agency for Science, Technology and Research (A*STAR), Singapore 138632, Singapore

[3] School of Information Science and Technology, Northwest University, Xi'an 710127, China

[4] School of Physics and Optoelectronics, Xiangtan University, Xiangtan 411105, China

[5] Institute of Brain-Inspired Intelligence, National Laboratory of Solid State Microstructures, School of Physics, Collaborative Innovation Center of Advanced Microstructures, Jiangsu Physical Science Research Center, Nanjing University, Nanjing 210008, China

[6] Quantum Innovation Centre (Q. InC), Agency for Science Technology and Research (A*STAR), Singapore 138634, Singapore

[7] Department of Mechanical Engineering, National University of Singapore, Singapore 117575, Singapore

[8] Department of Materials Science and Engineering, National University of Singapore, Singapore 117575, Singapore

[9] Department of Electrical and Computer Engineering, National University of Singapore, Singapore, 117583, Singapore

*Email: hanzhang@nwu.edu.cn, kahwee.ang@nus.edu.sg, yeesin_ang@sutd.edu.sg

Keywords: atomristor, 2D material; resistive switching, high-throughput simulation, Neuromorphic computing.

## Abstract

Atomristors, non-volatile resistive switching devices based on two-dimensional (2D) monolayers, are promising building blocks for energy-efficient memory and neuromorphic computing. However, their design remains restricted to a few materials such as $MoS_2$ and h-BN, limiting functional diversity and design flexibility. Here, a high-throughput computational framework combining density functional theory, machine-learning molecular dynamics, and quantum transport simulations screens about 2,900 exfoliable monolayers for vacancy-mediated resistive switching, identifying 17 thermally stable candidates in two mechanistically distinct classes. In Class 1 monolayers, such as GaS, Au adsorption at the native vacancy introduces conducting states, switching the insulating monolayer from a high- to a low-resistance state (HRS-to-LRS). Class 2 monolayers, comprising ionically bonded metal oxyhalides and nitrohalides such as BiOCl, exhibit previously unreported inverted switching. Vacancy-released electrons delocalize and push the Fermi level into the conduction band, placing the device natively in the LRS; Au adsorption re-localizes these carriers and returns the Fermi level to the gap, driving LRS-to-HRS switching. Quantum transport simulations confirm both mechanisms, while migration-barrier calculations identify the electrode-2D separation as a key parameter governing Au migration and the resistance window. These findings expand the atomristor landscape and establish complementary switching as a design paradigm for multifunctional memory and neuromorphic hardware.

## 1. Introduction

Two-dimensional (2D) materials are promising in revolutionizing memristor technology for nonvolatile data storage and neuromorphic computing.[1-8] The ultrathin nature of 2D metal-insulator-metal (MIM) devices overcomes the vertical scaling limitations faced by traditional oxide-based devices, paving the way for high-density, high-speed, and ultralow-power devices.[9,10] While 2D monolayers were initially considered unfeasible for memristor devices due to concerns about excessive leakage current,[11] such a view was changed in 2018 with the demonstration of nonvolatile resistive switching (NVRS) in monolayer transition metal dichalcogenides (TMDC) $MX_2$ (M = Mo, W; and X = S, Se).[12] These atomic-scale devices, named *atomristor*, can exhibit resistive switching (RS) through the adsorption of metal atoms from electrodes into existing native point vacancies. Metal adsorption introduces localized transport states near the Fermi level, abruptly converting the initially insulating monolayer into a conductive state and enabling ultrafast resistive switching.[13-20] Beyond memory applications, atomristors have also been employed as analog switches for high-frequency applications in emerging 5G and terahertz communication systems.[21] Recent successful integration of atomristors onto the back-end-of-line of silicon microchips further reveals their potential in next-generation computing electronics.[22]

Despite significant progress, the practical deployment of atomristors remains challenging due to low fabrication yield and poor endurance[23]. Addressing these challenges requires not only a deeper understanding of atomistic switching mechanisms but, more importantly, the systematic discovery of a broader and more diverse set of atomristor-compatible 2D monolayers. To date, atomristor research-both experimental and theoretical-has focused on a small subset of 2D materials, primarily a few transition-metal dichalcogenides[13,24,25] and hexagonal boron nitride (h-BN) [22,26,27]. Existing theoretical simulations have largely been localized to these specific materials, focusing on understanding the underlying RS mechanisms, metal-2D contact properties, and the dependence of switching voltage and ON/OFF resistance ratio on device parameters.[14,26,28-37] However, with the rapidly expanding library of 2D materials[38], there is a pressing need for a scalable computational workflow capable of identifying new atomristor candidates beyond the currently limited material pool.

In this work, we establish a high-throughput computational screening framework that integrates density functional theory (DFT), machine learning (ML)-accelerated molecular dynamics (MD), and quantum transport simulations to systematically explore atomristor functionality across a large pool of 2D monolayers. By assessing the stable vacancy defect configurations in 2D materials curated from the 2DMatPedia database[39], we examine how gold (Au), a widely used electrode material in RS devices [12,15,20], modifies the electronic properties of defective monolayers upon adsorption and enables atomic-scale RS. Our screening workflow uncovers two distinct functional classes. Class 1 monolayers exhibit the conventional insulating-to-metallic transition upon Au adsorption, corresponding to high resistance state (HRS) to low resistance state (LRS) switching. More importantly, we identify an unconventional Class 2 monolayers that display an *inverted* metallic-to-insulating (i.e., LRS-to-HRS) switching behaviour[40] that has not been previously reported in 2D materials. The adsorption of Au into a native vacancy defect site induces charge re-localization. Because these materials operate without an initial SET process, they are inherently suited for RESET-dominated workloads and may offer improved energy efficiency in circuit-level integration of atomristors.[40]

To bridge material discovery with device implementation, we assessed switching feasibility in a realistic Au-2D-Au vertical heterostructure by combining DFT-based nudged elastic band (NEB) calculations with large-scale MD simulations utilizing a benchmarked universal ML interatomic potential. This holistic approach ensures that candidate materials are rigorously assessed from their intrinsic electronic response to their kinetic stability in device-relevant configurations, providing clear guidance for experimental realization and application-specific material selection. Importantly, the discovery of Class 2 monolayers establishes an entirely new materials-level design space for atomristors by enabling intrinsically inverted switching at the atomic scale. Such complementary functionality could diversify the application scenario of atomristors, such as complementary resistive switches (CRS) for suppressing the sneak-path currents[41] in crossbar arrays, and CMOS-like complementary logic operation[42] as well as achieving better excitatory-inhibitory balance[43] in neuromorphic architectures.

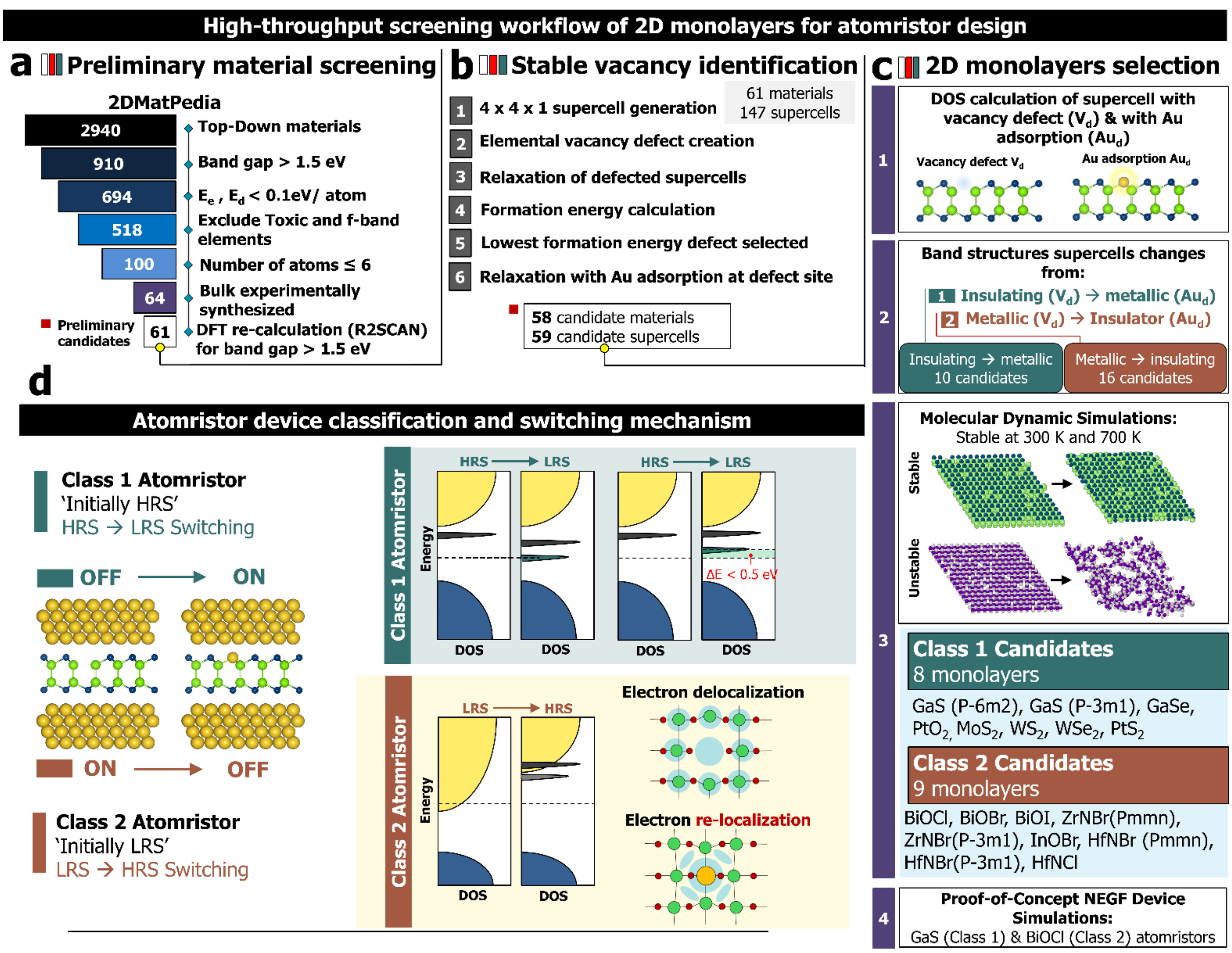

**Figure 1. Computational workflow for atomristor materials screening**. **(a)** Initial 2D materials screening based on the *2DMatPedia* Database. **(b)** Identification workflow of vacancy defects with the lowest formation energy as the most probable native defects of the monolayer supercells. **(c)** Calculations of the band structures and electronic density of states of supercells with a vacancy defect and with an Au atom adsorbed into the defect site. Two types of switching behaviours are identified, namely Class 1 (Class 2) candidates where the system changes from insulating (metallic) to metallic (insulating) upon gold atom adsorption into the vacancy defect site. After molecular dynamics stability checks, a final list of Class 1 and Class 2 monolayers is obtained. GaS and BiOCl are simulated using the DFT-NEGF method to assess their application as atomristors. **(d)** Atomristor devices are classified into Class 1 (HRS→LRS) and Class 2 (LRS→HRS) subclasses. Class 1 atomristor is the conventional atomristor subclass and is 'initially-HRS', i.e., the Class 1 atomristor is initially in the HRS when the native vacancy defect is present. Upon adsorption of an Au atom into the vacancy site, the creation of defect bands around the FL switches the Class 1 atomristor into LRS. In contrast, the Class 2 atomristor, a previously unknown atomristor subclass, is 'initially-LRS'. Class 2 atomristor is originally in the LRS and switches into HRS when the vacancy site is filled by an Au atom. Such an unconventional switching mechanism arises from electron delocalization caused by the vacancy defect, which raises the FL towards the conduction band, leading to LRS natively. Au atom adsorption localizes the electron and switches the device into HRS.

## 2. Results

### 2.1 Overview of computational screening and device modelling workflows

This work is structured into two primary phases: a high-throughput screening workflow to identify promising 2D materials for atomristors (**Figure 1**), followed by a rigorous analysis of the metal adsorption and desorption within a metal-2D-metal device architecture (**Figure 2**).

In the material screening workflow, starting with monolayer 2D materials from the 2DMatpedia database[39] (**Figure 1a**), we identify the most probable vacancy defect via formation energy calculations (**Figure 1b**). An Au atom is then introduced into the vacancy, and the electronic states of the pristine vacancy and Au-adsorbed structures are compared (**Figure 1c**). This screening reveals two functional classes (**Figure 1d**). Class 1: These monolayers start in an HRS, as their vacancy configuration is semiconducting or insulating. When Au atoms are adsorbed, they become metallic or semi-metallic, transitioning to an LRS, exhibiting the conventional RS behaviour. (**Figure 1d**). Class 2: These monolayers begin in LRS due to their metallic vacancy configuration, in which electrons are delocalized across the lattice, creating conductive pathways (**Figure 1d**). Upon Au adsorption, the electrons are re-localized, switching the monolayers into insulating HRS. Following extensive stability analysis at both room and high temperatures via molecular dynamics (MD) simulations, we identify 8 and 9 stable Class 1 and Class 2 monolayers, respectively. The resistive switching behavior is further validated using quantum transport simulations.

Following the screening, we assess the feasibility of Au adsorption and desorption in the identified materials (**Figure 2**). Migration barriers for Au transfer between the electrode and vacancy sites are calculated using DFT-based nudged elastic band (NEB) simulations (**Figure 2b**), and the results are corroborated with room-temperature MD simulations employing a universal machine-learning interatomic force field (**Figure 2b**). To ensure efficiency and accuracy, the ML potential is carefully chosen by benchmarking the top candidates from the MatBench[44] Discovery platform against DFT across four distinct energy and geometry metrics for the selected materials (**Figure 2a)**. In MD analysis, adsorption is classified as spontaneous when an Au atom migrates into the vacancy during the simulation trajectory; otherwise, the process is considered non-spontaneous (**Figure 2b**). Similarly, spontaneous desorption is defined by the migration of an Au atom initially occupying the vacancy back to the electrode within the MD time window (**Figure 2b**), indicating a volatile 'ON' state. By correlating static NEB barriers with dynamic MD trajectories, we evaluated the reversibility and thermal robustness of Au-mediated switching, thereby confirming the suitability of these materials for room-temperature resistive switching applications.

A comprehensive breakdown of the screening process and outcomes is presented below.

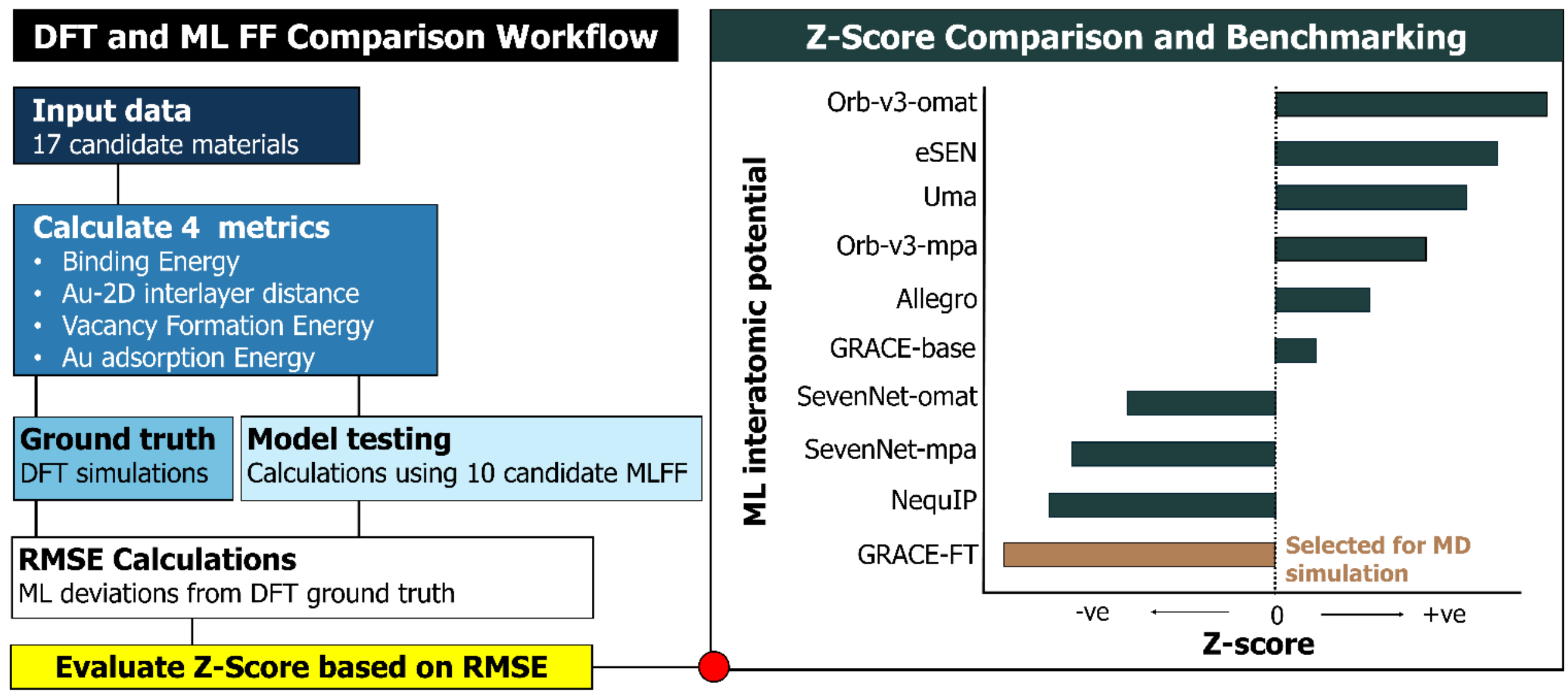

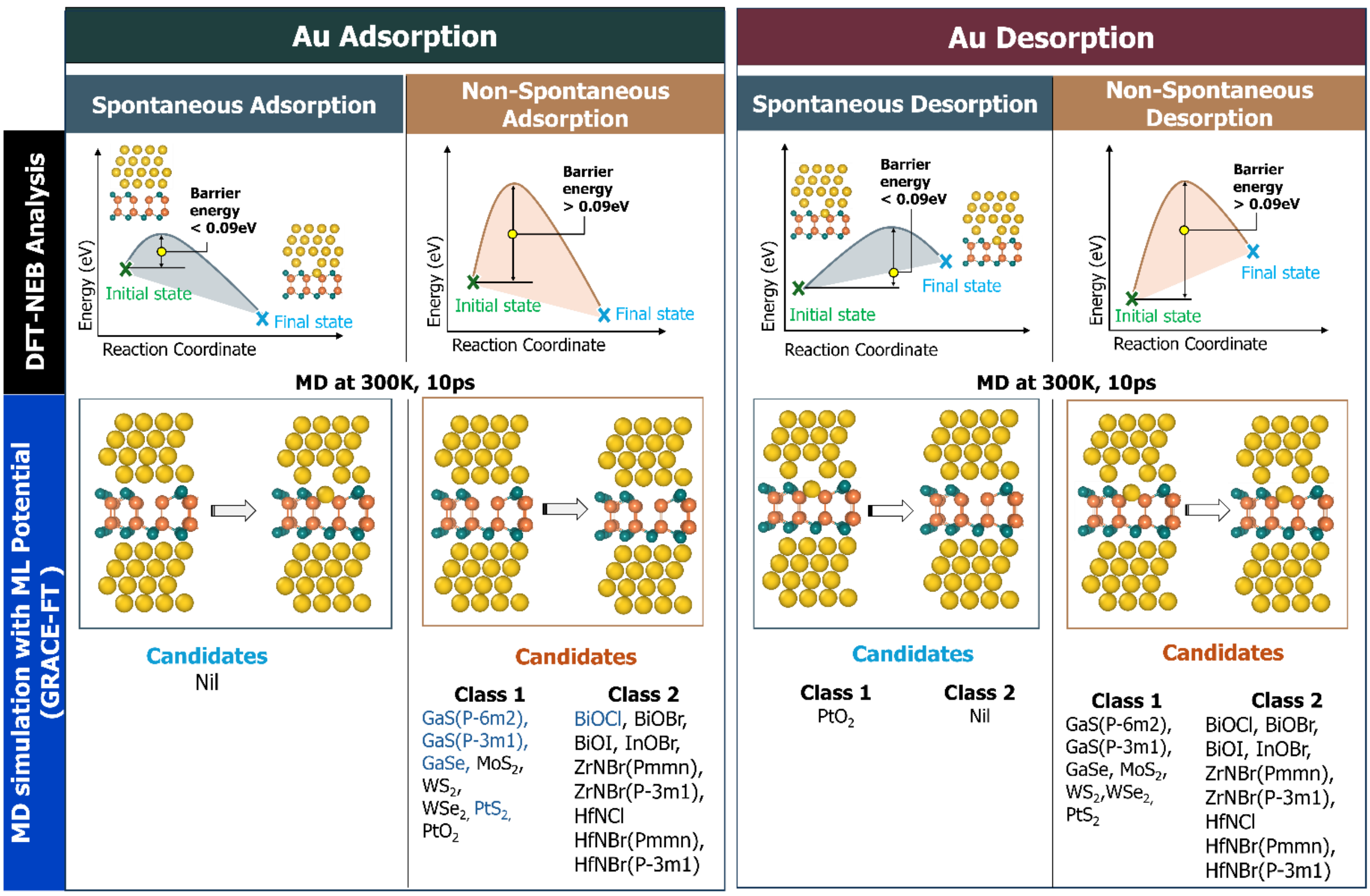

**Figure 2. Schematic of Au adsorption and desorption kinetics in an Au-2D-Au atomristor device. (a)** Workflow for benchmarking universal machine-learning (ML) interatomic potentials against DFT reference data. Based on the best overall agreement across 17 candidate materials, the Grace-FT potential is selected for subsequent molecular dynamics (MD) simulations. (b) Schematic of the Au adsorption and desorption protocols. DFT-based nudged elastic band (NEB) calculations are first used to determine energy barriers for Au migration from the electrode to a vacancy site (forward barrier) and from the vacancy back to the electrode (reverse barrier). These barriers are then correlated with finite-temperature MD simulations of a representative Au-2D-Au model performed at 300 K for 10 ps. For adsorption, a vacancy is introduced in the 2D layer, and spontaneous adsorption is identified when an Au atom migrates into the vacancy during MD; otherwise, adsorption is classified as nonspontaneous. For desorption, the vacancy is initially occupied by an Au atom, and spontaneous desorption is defined by Au migration back to the electrode during MD. Comparison of NEB and MD results reveals that spontaneous migration occurs when the NEB barrier is below ~0.09 eV. While Au adsorption is nonspontaneous for most candidates, spontaneous adsorption is observed for four Class 1 and one Class 2 candidates (marked in blue) at equilibrium electrode-2D separation and can be suppressed by increasing the interfacial distance. In contrast, one Class 1 material ($PtO_2$) exhibits spontaneous Au desorption even at increased interface separation.

## 2.2 2D material screening

The monolayer screening is based on the 2DMatPedia[39] database. Initially, we select 2,940 "top-down" materials that are theoretically exfoliable from layered bulk parents listed in the Materials Project (MP) database[45,46]. Since only insulating materials are suitable as active layers in atomristor, we filter for materials with a band gap greater than 1.5 eV, narrowing the list to 910 candidates. To ensure both exfoliation potential and energetic stability, we further apply a cutoff of 100 meV/atom for both exfoliation and decomposition energies, resulting in 694 materials. Hazardous elements (Pb, As, Hg, Cr, Cd, Be, U, Th) are excluded to minimize health and environmental risks. Due to the complexity of handling *f*-electrons in DFT, materials containing *f*-block elements (Er, Ce, Pr, Nd, Pm, Sm, Eu, Gd, Tb, Dy, Ho, Tm, Yb, Lu) are further removed. As defect formation simulations require supercells, we limit the unit cell size to six atoms or fewer to manage computational costs. Lastly, we verify the experimental availability of the corresponding bulk materials in the MP database (**Figure 1a**). The above criteria yield 64 monolayers from the 2DMatPedia database. Since our study employed the META-GGA functional (r2SCAN)[47], which differs from the vdW-optB88 exchange-correlation functional used in developing 2DMatPedia, we recalculate the electronic band structures. 61 of the 64 materials remain within the initial band gap threshold of >1.5 eV. These candidates include metal halides, metal chalcogenides, metal oxides, metal oxyhalides, metal nitrohalides and metal hydroxides (see a list in **Supplementary Table 1**).

## 2.3 Defect formation in 2D monolayers

Vacancy defects are most commonly associated with NVRS in atomristors. Both computational studies and atomic-scale scanning tunneling microscopy (STM) have shown that Au atoms from electrodes can adsorb into sulfur vacancies in $MoS_2$, triggering a transition from insulating to metallic behavior[15,29,32]. Since such vacancies often emerge naturally during fabrication, a systematic understanding of native point defects is essential for developing reliable atomristor devices. Here we introduce all possible elemental monovacancies into the 61 screened materials, generating 147 defective supercells (**Figure 1b**). For each supercell, we calculate its vacancy formation energies (see **Methods**) and identify the lowest-energy one as the native defect configuration.

We observe the following trends across different chemical families. In metal monohalides (e.g. AgI, AuI, CuBr), metal (M) vacancies dominate; for instance, the silver vacancy ($V_{Ag}$) in AgI is more stable than the iodine vacancy ($V_I$). In contrast, metal dihalides typically favour halogen (X) vacancies, with $GeI_2$ as an exception, where Ge vacancies are more favourable. This is because $GeI_2$ has more covalent Ge-I bonds (**Supplementary Figure 1a**); after removing the Ge atom, the surrounding I atoms can relax to form stable configurations, offsetting the energy cost. However, other dihalides like $CaI_2$ exhibit predominantly ionic bonding (**Supplementary Figure 1b**), making I vacancies more favourable due to the lower energy cost of removing a singly charged anion ($I^-$) rather than a doubly charged cation ($Ca^{+2}$). The lowest-energy vacancies for all monolayers are shown in **Supplementary Table 2**.

In chalcogenides (sulfides, selenides, tellurides), chalcogen vacancies dominate, which is consistent with previous studies[15]. However, SnS is an exception where both Sn and S vacancies exhibit nearly identical formation energies. This behavior stems from the orthorhombic Pmmn structure of SnS, which lacks direct Sn-Sn bonding across planes. In contrast, materials like GaS, GaSe, and InSe (typically in hexagonal P-6m2 or P-3m1 structures) have vertically aligned metal atoms forming metal-metal covalent bonds. Bader charge analysis (**Supplementary Figure 1c**) confirms both strong metal-chalcogen charge transfer and additional interlayer metal-metal interactions. Thus, removing a metal atom disrupts multiple strong covalent bonds, making chalcogen vacancies more favorable. In SnS, where bonding is more evenly distributed between Sn and S atoms (**Supplementary Figure 1d**), removing either atom leads to similar energetic penalties. In h-BN, the nitrogen vacancy shows a lower formation energy than the boron vacancy in a neutral environment. In metal oxides, oxygen vacancies are more stable than metal vacancies. However, in metal oxyhalides and metal nitrohalides, halogen vacancies dominate due to weaker M-X bonding than M-O(/N) bonding (**Supplementary Figure 1e**). In metal hydroxides, hydrogen vacancies are the most favourable energetically.

The near-zero or even negative formation energies for Cu vacancies in Cu-based monohalides suggest spontaneous vacancy formation. While the presence of vacancies is essential for enabling the RS behavior, an uncontrolled increase in vacancy concentration may lead to reduced structural stability and limited device reliability. Consequently, we exclude such monolayers from further screening. From the remaining 58 monolayers, we select the most favourable vacancy for each monolayer while retaining both Sn and S vacancies in SnS, thus leading to 59 defective supercells for further investigation in the next stage.

### 2.4 Electronic states of defects and atomristor classifications

Electronic structure phase transition (e.g., insulator-to-metal transition) induced by the adsorption of a metal atom into the vacancy defect site is essential to achieve the RS functionality of an atomristor. To capture this, we compute the electron density of states (DOS) of both vacancy and Au-adsorbed supercells (**Figure 1c**). Here, we chose to study Au as Au electrodes are prevalent in 2D RS devices.[12,15,20] To systematically track adsorption-induced transitions across a large material set, we evaluate the effective band gap for both vacancy and Au-adsorbed structures. We define the effective band gap as the energy difference between the highest occupied and the lowest unoccupied electron state, explicitly including defect-induced localized states.

We identify an initial pool of 26 candidates exhibiting electronic phase transitions required for RS. The candidates are classified into two distinct classes, namely: (i) Class 1 monolayers undergoing an insulator-to-metal transition; and an unconventional family of (ii) Class 2 monolayers that undergo the *inverse switching* of a metal-to-insulator transition upon Au adsorption. The final candidates of Class 1 and Class 2 monolayers are summarized in **Figures 3a and 3b**, respectively. Their band gap and band alignment are shown in **Figure 3c**. In the following sections, we detail the screening workflow and the switching mechanisms of the Class 1 and Class 2 monolayers.

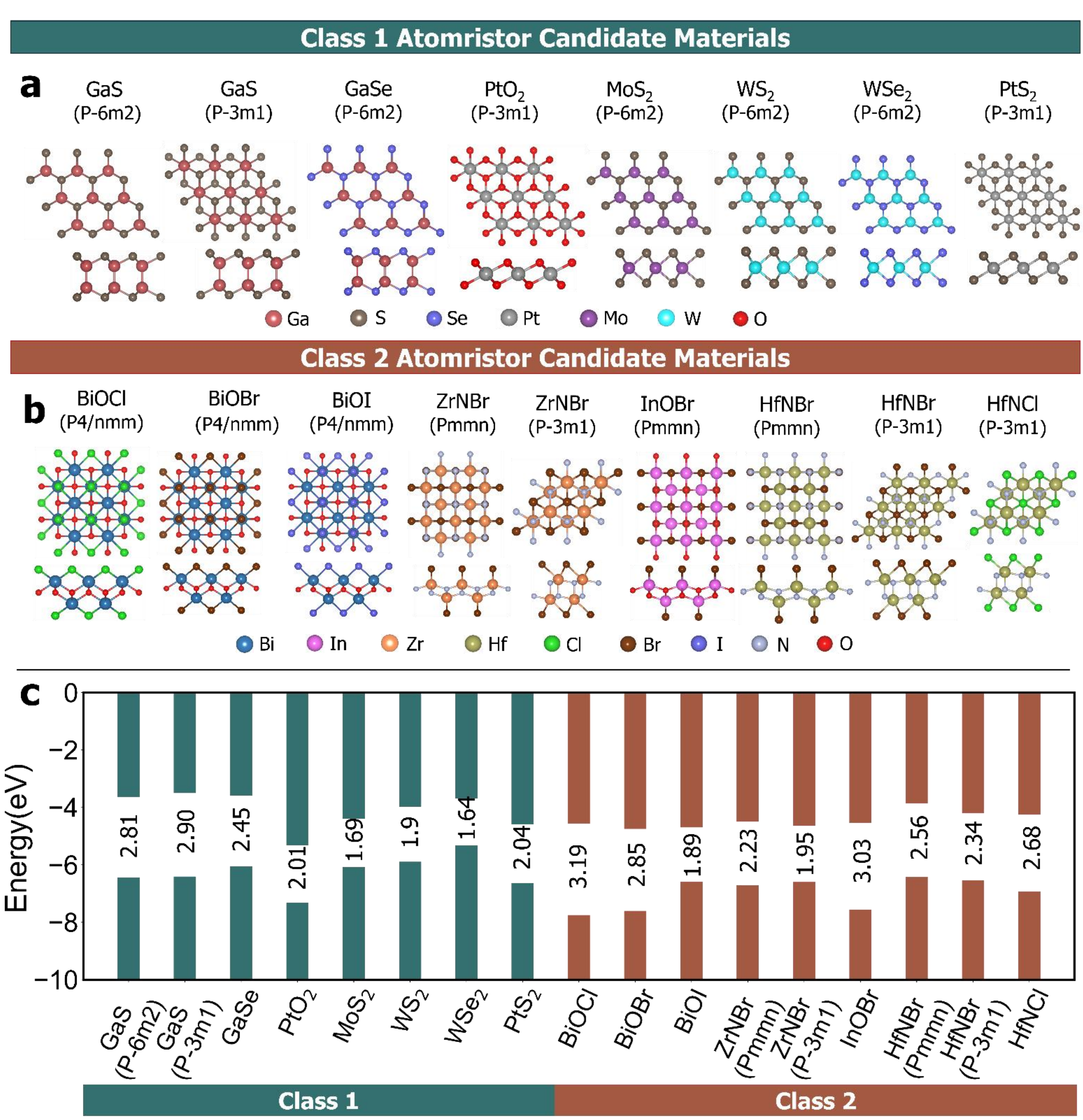

**Figure 3. Structural and Electronic Details of Final Atomristor Candidates. (a)** Top-view and side-view images of atomic structures of Final 8 Class 1 candidates, **(b)** top-view and side-view images of atomic structures of final 9 Class 2 candidates. Each material is identified by its name and space group. **(c)** Electronic band alignments for Class 1 (green bars) and Class 2 (red bars). Band gap values (in eV) are indicated for each material. Materials with identical elemental compositions (e.g., GaS) are differentiated by their space group, noted in braces.

***Class 1 Monolayers:*** The vacancy-defective supercell starts as semiconducting or insulating but becomes metallic or with a significantly reduced effective band gap after Au adsorption, reflecting the typical HRS-to-LRS transition in atomristors[12,20,22]. To identify such monolayers, the screening criterion requires the initial vacancy state to have an effective band gap larger than 1 eV, while that of the Au-adsorbed configuration becomes lower than 0.5 eV. The > 1 eV cutoff ensures the defective material remains reliably insulating in the initial HRS. In contrast, the < 0.5 eV threshold guarantees that Au adsorption can effectively drive the system into a metallic or strongly doped regime with significantly lowered resistance (i.e., LRS). Monolayers that satisfy these conditions are predominantly chalcogenides and oxides. In total, 10 candidates are identified: (i) 8 metal chalcogenides (GaS-P6m2, GaS-P3m1, GaSe, InSe, $MoS_2$, $WS_2$, $WSe_2$, $PtS_2$), (ii) 1 oxide ($PtO_2$), and (iii) 1 mixed-anion compound (GaTeCl).

For instance, in GaS (P-6m2), removing an S atom generates two defect states, one below the CBM and another above the VBM (**Figure 4a**). After Au adsorption, these states shift and intersect the FL, inducing metallicity in the monolayer (**Figure 4a**). Similar behaviours are also observed in other chalcogenides GaSe, InSe, $PtS_2$, $MoS_2$, $WS_2$ (**Figure 4b**), $WSe_2$, as well as oxide $PtO_2$ (see **Supplementary Figure 2a**). In GaTeCl, the localized states formed due to a Te vacancy result in an effective band gap of 1.36 eV, while Au adsorption causes these bands to shift closer to the FL, reducing the effective band gap to 0.44eV (see **Supplementary Figure 2b**), thereby satisfying the screening criteria defined for Class 1 monolayers.

***Class 2 Monolayers:*** Class 2 monolayers are characterized by a vacancy-induced conductive electronic state that transforms into an insulating state following Au adsorption. We identify Class 2 monolayers by applying a screening criterion where the parent vacancy structure has an effective band gap less than 0.5 eV, while Au adsorption increases the gap above 1.0 eV. Class 2 monolayers are predominantly halogen-containing materials, such as metal monohalides, dihalides, oxyhalides, and nitrohalides.

In metal monohalides, removing a metal atom (lowest-energy vacancy) typically shifts the FL into the valence band, generating strong p-type characteristics (see **Supplementary Figure 3a**). This happens because the absence of a positively charged metal center creates an acceptor-like defect. Remarkably, when this vacancy is occupied by an Au atom, the band gap reopens, causing a sharp increase in resistance and signifying an LRS-to-HRS switching (**Supplementary Figure 3a**). Six monohalides are obtained: AgI (P3m1), AgI (P4/nmm), AuI, AuBr, TlF, and TlBr. The microscopic origin of this band-gap reopening is discussed in the following sections.

In the metal dihalide family, halogen vacancies typically create in-gap states with effective band gaps > 0.5 eV and thus fall outside our criterion (see **Supplementary Figure 3b**). However, $MnBr_2$ represents a notable exception; its Br vacancy pushes the FL into the conduction band, resulting in an n-type LRS (**Supplementary Figure 3c**). Au adsorption at the vacancy site re-establishes the band gap, reverting $MnBr_2$ into HRS.

Within the metal oxyhalide and nitrohalide families, 9 monolayers exhibit LRS-to-HRS switching upon Au adsorption at the halogen vacancy sites. For instance, BiOCl (**Figure 4c**), BiOBr, BiOI, and ZrNBr (Pmmn) show a pronounced metallic character upon halogen vacancy formation, as the FL shifts into the conduction band. Meanwhile, other members of this family, i.e., InOBr, HfNBr (Pmmn), HfNBrN (P-3m1), HfNCl, and ZrNBr (P-3m1), are initially in LRS, as the halogen vacancies bring the Fermi level closer to the conduction band with an effective band gap < 0.5 eV **(Figure 4d).** In all 9 materials, Au adsorption establishes a sizable band gap > 1 eV, thus indicating the unconventional LRS-to-HRS switching.

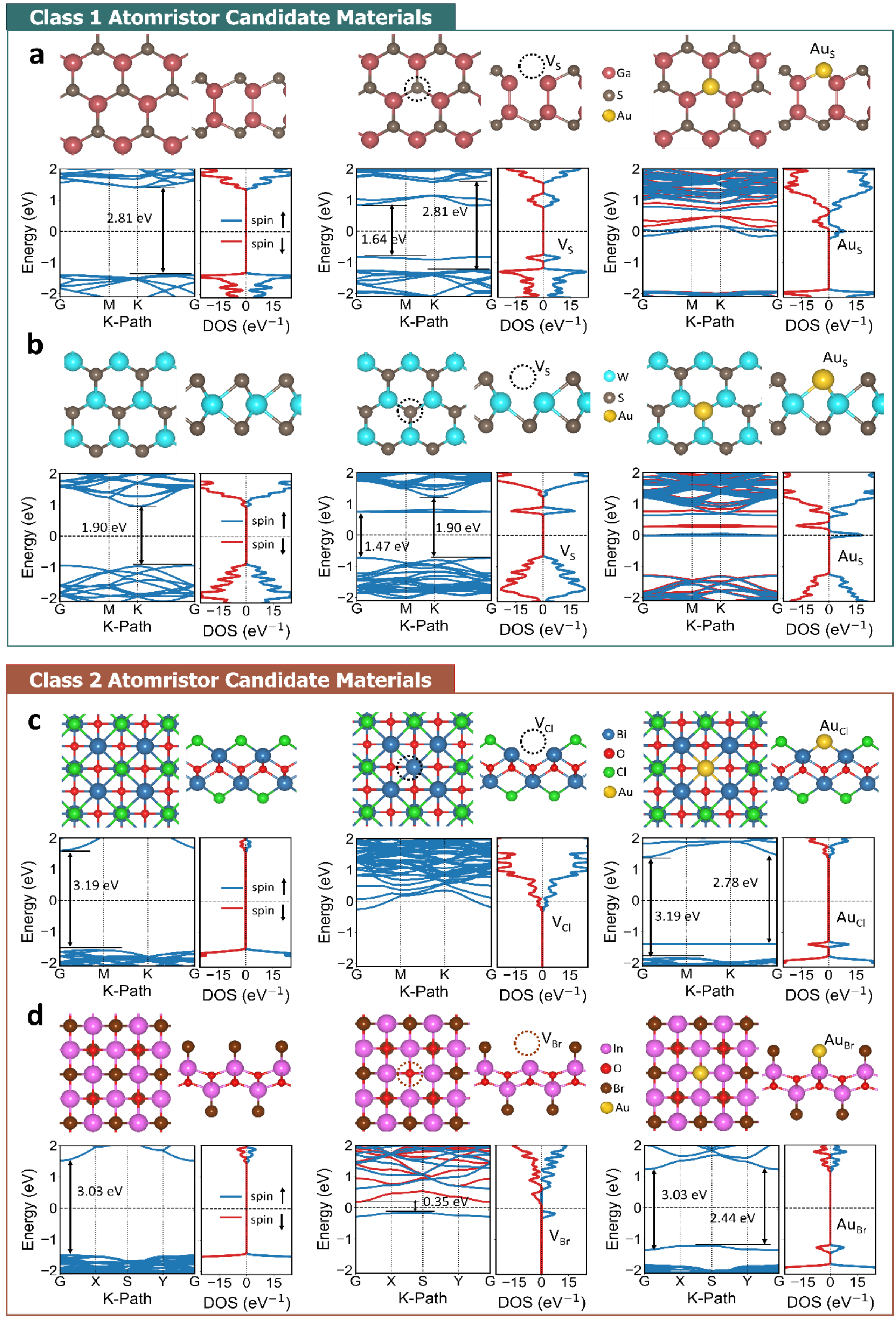

**Figure 4. Evolution of Electronic Band Structures in Class 1 and Class 2 Monolayers.** Band structures for pristine (left) to vacancy (middle) and Au-adsorbed vacancy (right) states for **(a)** GaS (P-6m2), **(b)** $WS_2$, **(c)** BiOCl and **(d)** InOBr. The dotted black circles represent the most probable vacancies present in the materials (S vacancy in $WS_2$ and GaS, Cl vacancy in BiOCl, and Br vacancy in InOBr). The FL is set at 0 eV. All pristine materials are semiconducting or insulating. In Class 1 monolayers, S vacancies introduce localized in-gap states, and Au adsorption further modifies the electronic structure by shifting the FL (GaS) or creating new states at it ($WS_2$). In Class 2 monolayers, the Cl vacancy in BiOCl drives a metallic state by pushing the FL into the conduction band, while the Br vacancy in InOBr brings the conduction band extremely close to the FL, generating a quasi-metallic state. Remarkably, for both Class 2 monolayers, Au adsorption at the vacancy site re-establishes a finite band gap, successfully restoring their insulating properties.

In total, we identify 16 Class 2 candidates that exhibit the unconventional LRS-to-HRS switching upon Au adsorption. In **Table 1**, all 26 promising Class 1 and Class 2 candidates, along with their electronic properties, are listed. For completeness, the effective band gaps of the vacancy and Au-adsorbed supercells of the remaining 32 materials (excluded by the screening workflow) are also provided in **Supplementary Table 4**.

### 2.5 Native Defect stability assessment

The thermal stability assessment of the candidates with native defects is also integrated in our screening workflow (**Figure 1c**). We perform MD simulations centred on the most probable vacancy configurations of the 26 candidates obtained thus far (i.e., 10 Class 1 and 16 Class 2 monolayers). Given the computational demands of *ab initio* MD, we employed the universal machine learning force field, Orb-v3[48], which enables efficient and reliable simulation of defect behaviour under thermal stress. Simulations are performed at room temperature (300 K) and at elevated temperature (700 K) to mimic device operation conditions, where prolonged electric fields can cause localized heating.

Among the Class 1 monolayers, GaTeCl exhibits instability upon the introduction of Te vacancies, despite the low formation energy of this defect. The neighbouring Cl atoms migrated into the Te vacancy site, causing significant structural rearrangement (see **Supplementary Figure 4a**). As a result, GaTeCl is excluded from further consideration. To simulate realistic defect scenarios in our MD simulations, we employ an automated code to randomly introduce vacancies, which sometimes results in double vacancies in the materials. In InSe, such double-vacancy regions showed local structural distortion at high temperatures (see **Supplementary Figure 4b**). To ensure the robust stability of the defective systems, we also removed InSe from our material pool. After performing the MD stability screening, we identified 8 high-stability Class 1 monolayers (**Table 1**) capable of HRS-to-LRS switching (see **Supplementary Figure 4c**) across varied temperatures.

Among the Class 2 monolayers, all 6 metal monohalides show structural degradation at elevated temperature when metal vacancies are present (see **Supplementary Figure 5a**). In addition, the $MnBr_2$ structure with Br vacancies is distorted at room temperature (see **Supplementary Figure 5b**). Excluding such 7 unstable monolayers (see **Supplementary Figure 5c**), we obtain 9 Class 2 monolayers (**Table 1**) that retain structural stability at both 300 K and 700 K thermal conditions.

**Table 1. Summary of Key Properties for Atomristor Candidates.** This table presents the structural and electronic properties of promising atomristor candidates. It includes the space group, initial bandgap, the type of vacancy introduced, the resulting bandgap after vacancy formation, the bandgap after Au adsorption, the determined resistive switching (RS) class (Class 1 or Class 2), and the MD-calculated defect stability. Band gaps for defective and Au-adsorbed systems are estimated based on the energy difference between the highest energy state in the valence band and the lowest energy state in the conduction band, including any localized defect states. For instance, in GaS (P-6m2), the bandgap for the vacancy state is measured between the two localized states found just below CBM and above the VBM, as illustrated in Figure 3a. Materials marked as RS Class ++ exhibit a quasi-metallic state within either Class 1 or Class 2. For instance, in GaTeCl after Au adsorption, or in InOBr following a Br vacancy (Figure 3d), the electronic states appear very close to the FL but do not directly cross it, indicative of this quasi-metallic nature. Materials marked with * indicate final candidates selected based on both switching behavior and defect stability. Materials marked with + have been previously reported in experimental atomristor devices. The unit of bandgap is eV.

| Material | Space-group | Bandgap (Pristine) | Vacancy Type | Bandgap (Vacancy) | Bandgap (Au adsorbed) | RS Class | Defect Stability |
|---|---|---|---|---|---|---|---|
| $GaS^{*}$ | P-6m2 | 2.81 | S | 1.64 | 0.00 | 1 | Yes |
| $GaS^{*}$ | P-3m1 | 2.90 | S | 1.56 | 0.00 | 1 | Yes |
| $GaSe^{*}$ | P-6m2 | 2.45 | Se | 1.36 | 0.00 | 1 | Yes |
| InSe | P-6m2 | 1.96 | Se | 1.34 | 0.00 | 1 | No |
| $PtO_2^{*}$ | P-3m1 | 2.01 | O | 1.30 | 0.00 | 1 | Yes |
| $MoS_2^{*+}$ | P-6m2 | 1.69 | S | 1.21 | 0.00 | 1 | Yes |
| $WS_2^{*+}$ | P-6m2 | 1.90 | S | 1.47 | 0.00 | 1 | Yes |
| $WSe_2^{*+}$ | P-6m2 | 1.64 | Se | 1.34 | 0.00 | 1 | Yes |
| $PtS_2^{*+}$ | P-6m2 | 2.04 | S | 1.00 | 0.00 | 1 | Yes |
| GaTeCl | Pmn2_1 | 2.71 | Te | 1.36 | 0.44 | $1^{++}$ | No |
| AgI | P-3m1 | 2.50 | Ag | 0.00 | 2.42 | 2 | No |
| AgI | P4/nmm | 2.56 | Ag | 0.00 | 2.43 | 2 | No |
| AuBr | Cmme | 2.40 | Au | 0.00 | 2.40 | 2 | No |
| AuI | Cmme | 2.11 | Au | 0.00 | 2.11 | 2 | No |
| TlBr | P4/nmm | 3.50 | Tl | 0.00 | 3.23 | 2 | No |
| TlF | P4/nmm | 4.08 | Tl | 0.00 | 2.14 | 2 | No |
| $MnBr_2$ | P-3m1 | 2.67 | Br | 0.00 | 2.18 | 2 | No |
| $BiOCl^{*}$ | P4/nmm | 3.19 | Cl | 0.00 | 2.78 | 2 | Yes |
| $BiOBr^{*}$ | P4/nmm | 2.85 | Br | 0.00 | 1.97 | 2 | Yes |
| $BiOI^{*}$ | P4/nmm | 1.89 | I | 0.00 | 2.00 | 2 | Yes |
| $ZrNBr^{*}$ | Pmmn | 2.23 | Br | 0.00 | 2.23 | 2 | Yes |
| $ZrNBr^{*}$ | P-3m1 | 1.95 | Br | 0.13 | 1.83 | $2^{++}$ | Yes |
| $InOBr^{*}$ | Pmmn | 3.03 | Br | 0.35 | 2.45 | $2^{++}$ | Yes |
| $HfNBr^{*}$ | Pmmn | 2.56 | Br | 0.49 | 2.54 | $2^{++}$ | Yes |
| $HfNBr^{*}$ | P-3m1 | 2.34 | Br | 0.39 | 2.22 | $2^{++}$ | Yes |
| $HfNCl^{*}$ | P-3m1 | 2.68 | Cl | 0.46 | 2.46 | $2^{++}$ | Yes |

### 2.6 Forward and inverse switching mechanisms

To understand the underlying mechanistic differences between Class 1 and Class 2 monolayers, we analyse the charge transfer and distribution in two representative candidates: GaS (Class1) and BiOCl (Class 2), before and after Au adsorption. In Class 1 monolayers, the bonds are primarily covalent. The charge density plot of GaS shows a substantial accumulation of electron density along the bonds connecting the Ga and S atoms (**Figure 5a**). When such a covalently bonded atom is removed, its neighbouring atoms form dangling bonds. Bader charge analysis reveals that when an S atom is removed, the electrons from the missing atom become localized on the neighbouring Ga atoms (**Figure 5b**), creating localized states within the band gap (**Figure 4a**). Further band-decomposed charge density analysis confirmed that these localized defect states, found both below the conduction band and above the valence band, arise from electron localization around the S vacancy (**Figures 5d, 5e**).

When an Au atom adsorbs into the vacancy, it attracts some of these localized electrons. The partial charge density plot confirms that the bands appearing around the FL are primarily generated from the electrons surrounding the adsorbed Au atoms (**Figure 5f**). However, Bader charge analysis shows that although electrons are transferred to the Au atom, it does not gain the same negative charge as the original S atom it replaces (**Figure 5c**). In pristine GaS, a typical S atom has a charge of about -0.86 e, whereas the charge on an Au atom at an S vacancy is only about -0.51 e. This reduction arises from Au's lower electronegativity and different valence electron configuration compared to S. Because the Au atom cannot fully restore the missing electrons of the removed S atom, excess electrons redistribute among the nearest Ga atoms, with noticeable changes also in the next-nearest Ga neighbours (**Figure 5c**). The incomplete charge re-localization at the Au site electron-dopes the system and shifts the FL towards the conduction band, thus leading to an insulator-to-metal transition.

Contrary to Class 1, Class 2 materials generally exhibit ionic bonding. The charge density plot of BiOCl shows electrons largely localized around the Cl atoms rather than shared between the Bi and Cl bonds (**Figure 5g**). However, ionicity alone is insufficient to produce Class 2 behavior. A crucial additional requirement is that the electrons released by vacancy formation can delocalize into extended conduction-band states rather than remaining trapped in localized defect levels. Bader charge analysis reveals that when a Cl vacancy forms, electrons from the missing atom are not confined to its nearest neighbours. Instead, the excess electrons delocalize, diffusing throughout the material and making even distant Bi atoms negatively charged (**Figure 5h**). Consequently, the system becomes electron-doped and behaves like a heavily n-type doped semiconductor, with the conduction band shifting towards the FL (**Figure 4c**). Partial (band-decomposed) charge density (**Figure 5j**) of the conduction band states around the FL shows that while electron density is the highest around the vacancy site, substantial spreading to distant Bi atoms is also present, thus further suggesting a substantial extent of electron delocalization in the vacancy-defective supercell.

Upon Au adsorption, previously delocalized electrons become *re-localized* around the Au atom. Bader charge analysis shows that the Au atom acquires a charge of approximately $-0.48$ e, as compared to $-0.69$ e is typically found on a Cl atom in the pristine material (**Figure 5i**). Hence, not all electrons originally associated with the Cl site are transferred to Au; some remain on the nearest Bi atoms. More distant Bi atoms, however, recover their pristine charge states as shown in the partial difference charge density plot in **Figure 5k**, indicating that the electrons are re-localized around the adsorption site. Such electron re-localization effect reverses the electron doping effect of the initial vacancy-defective configuration and shifts the FL back towards the mid-gap location, thus leading to LRS-to-HRS switching that is opposite to the case of Class 1 monolayers.

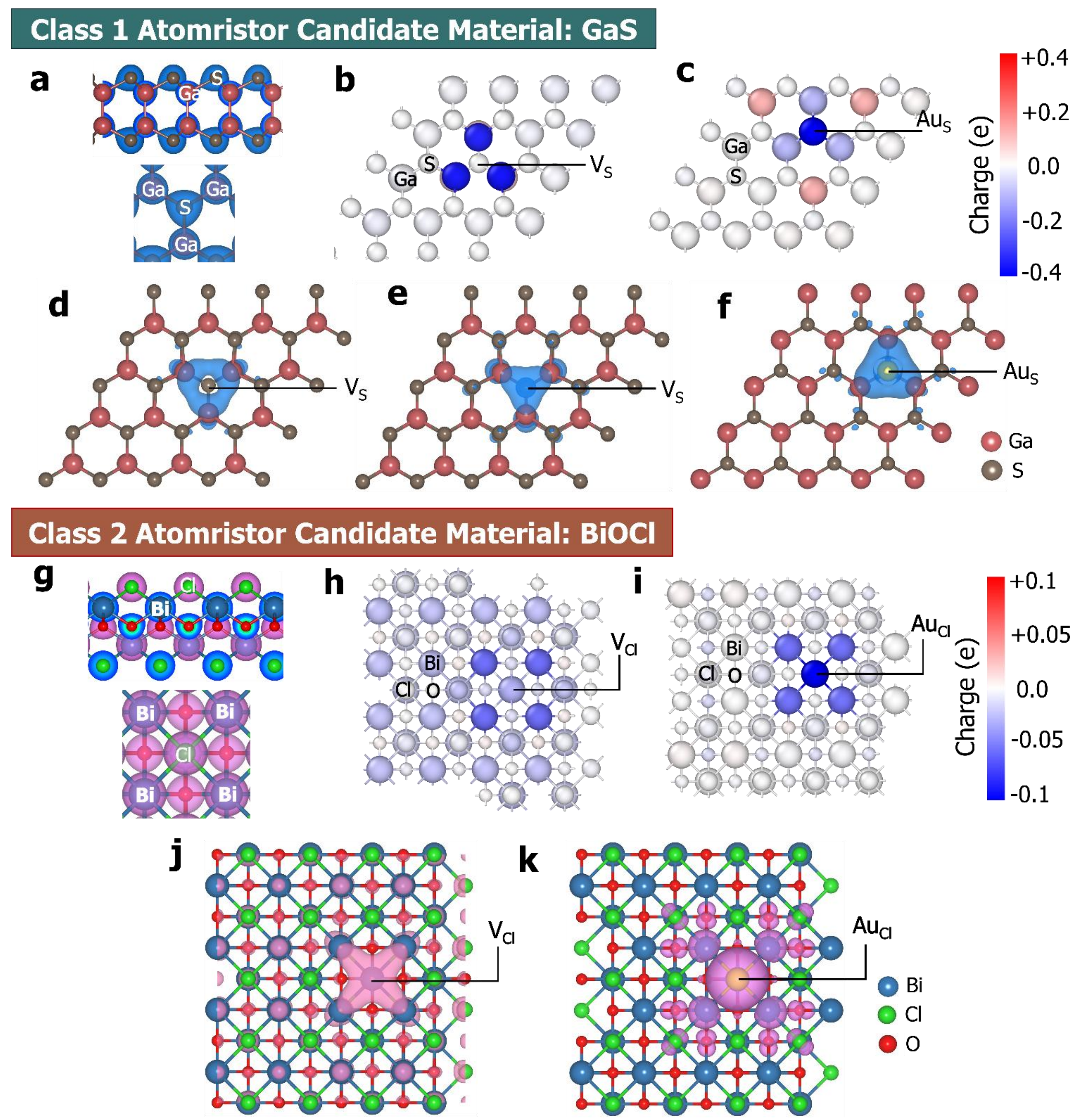

**Figure 5. Charge analysis of Class 1 and Class 2 materials. (a)** Side (top) and top (bottom) views of the charge density isosurface (0.075 eÅ$^{-1}$) in pristine GaS, highlighting S-Ga bonding. **(b, c)** Bader charge analysis of GaS containing an S vacancy ($V_S$) and with adsorbed Au ($Au_S$), respectively. Colors on the atoms represent the change in their Bader charge relative to pristine GaS; negative values denote electron gain and positive values denote electron loss. For Au adsorption, the reported charge for the Au atom is its total Bader charge, since it is absent in the pristine structure. For visual clarity, Ga atoms are drawn larger than S. **(d, e)** Partial charge density isosurfaces (0.003 eÅ$^{-1}$) of localized defect bands of $V_S$-GaS (below the conduction band and above the valence band, respectively). **(f)** Partial charge density isosurface (0.003 eÅ$^{-1}$) of states near the FL (±0.2 eV) in $Au_S$-GaS. **(g)** Side (top) and top (bottom) views of the charge density isosurface (0.075 eÅ$^{-1}$) in pristine BiOCl, illustrating Cl-Bi bonding. **(h, i)** Bader charge analysis of BiOCl with a Cl vacancy ($V_{Cl}$) and with adsorbed gold ($Au_{Cl}$), respectively. Atom colors represent the difference in atomic Bader charges relative to pristine BiOCl. For Au adsorption, the reported charge for the Au atom is its total Bader charge. For visual representation, Bi, Cl, and O are drawn in decreasing size order. **(j)** Partial charge density isosurface (0.001 eÅ$^{-1}$) of states near the FL (±0.2 eV) in $V_{Cl}$-BiOCl, **(k)** Partial charge density isosurface (0.001 eÅ$^{-1}$) of the localized state located inside the band gap above the valence band in $Au_{Cl}$-BiOCl. In GaS and BiOCl, the charge density isosurfaces are depicted in blue and pink, respectively.

## 2.7 Quantum transport simulations

To validate that defect-driven modulations of the electronic structure can indeed translate into measurable resistance changes under realistic device conditions, we perform quantum transport simulations (**Figure 6**) using a vertical Au-2D-Au atomristor geometry. We consider two representative monolayer candidates: GaS (Class 1) and BiOCl (Class 2) to illustrate the conventional and inverted atomristor operations, respectively. In GaS, the HRS arises from an S vacancy, while Au adsorption within that vacancy produces the LRS (**Figure 6a**). Conversely, in BiOCl, the intrinsic Cl vacancy defines the LRS, while Au adsorption signifies the HRS (**Figure 6d**). We calculate the HRS/LRS resistance ratio, or the memory window, at a bias voltage of 0.1 V. We consider an idealized configuration in which a fixed Au-2D separation of 5 Å is imposed to minimize electrode-induced effects, ensuring that the calculated resistance ratios reflect intrinsic properties of the 2D monolayer rather than interfacial interference[13].

For the GaS atomristor, Au adsorption reduces the resistance by a factor of $3.8\times10^2$, corresponding to conventional HRS-to-LRS switching (**Figure 6b**). In contrast, BiOCl exhibits an increase in resistance of approximately $0.6\times10^2$ upon Au incorporation, consistent with inverted LRS-to-HRS switching (**Figure 6e**). Current-voltage characteristics are further calculated over the bias range of 0-0.6 V (**Figure 6c, 6f**). At 0.6 V (−0.6 V), the current ratio $I_{LRS}/I_{HRS}$ reaches ~$2\times10^2$ ($3.7\times10^2$) for GaS, and ~$2.4\times10^1$ ($8.4\times10^1$) for BiOCl, respectively. The pronounced contrast in transport behavior between the two functional classes provides compelling, device-level validation of the defect-driven electronic-structure trends identified in the screening analysis.

We further investigated the influence of interfacial proximity on the resistance states. As shown in the **Supplementary Figure 6**, comparisons at 0.1 V for three different Au-2D interface distances reveal a pronounced reduction in the memory window as the electrode approaches the 2D layer. This degradation is attributed to the emergence of metal-induced gap states (MIGS) in the 2D monolayers upon contact with metallic electrodes. Our band structure calculations (**Supplementary Figure 7**) show that substantial MIGS emerge in the GaS/Au and BiOCl/Au contact heterostructures at lower distances. These MIGS, arising from the interfacial interaction between the electrodes and the monolayer, can generate a mid-gap conduction pathway which lowers the HRS resistance, thus narrowing the memory window.

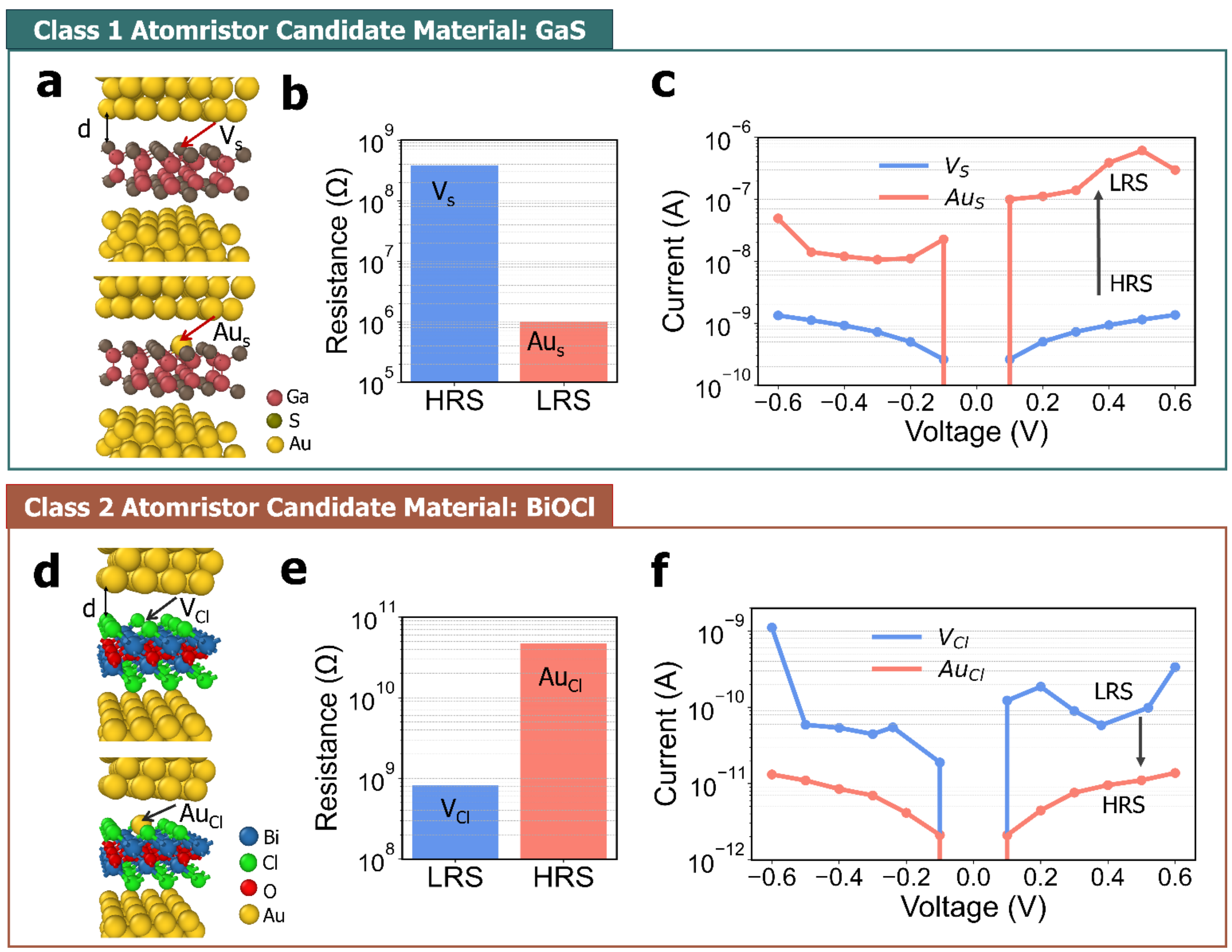

**Figure 6. Quantum transport simulations. (a-c)** Representative Class 1 atomristor, Au-GaS (P-6m2)-Au: **(a)** Device snapshots showing S-vacancy GaS in the HRS and Au-adsorbed GaS in the LRS sandwiched between top and bottom Au electrodes. **(b)** Calculated resistances of the S-vacancy and Au-adsorbed states at a bias of 0.1V. **(c)** Corresponding current-voltage (I-V) characteristics. **(d-f)** Representative Class 2 atomristor, Au-BiOCl-Au: **(d)** Device snapshots illustrating Cl-vacancy BiOCl in the LRS and Au-adsorbed BiOCl in the HRS**. (e)** Calculated resistances of the Cl-vacancy and Au-adsorbed states at 0.1 V. **(f)** Corresponding I-V characteristics. In all simulations, a fixed metal-2D monolayer separation of 5Å is employed.

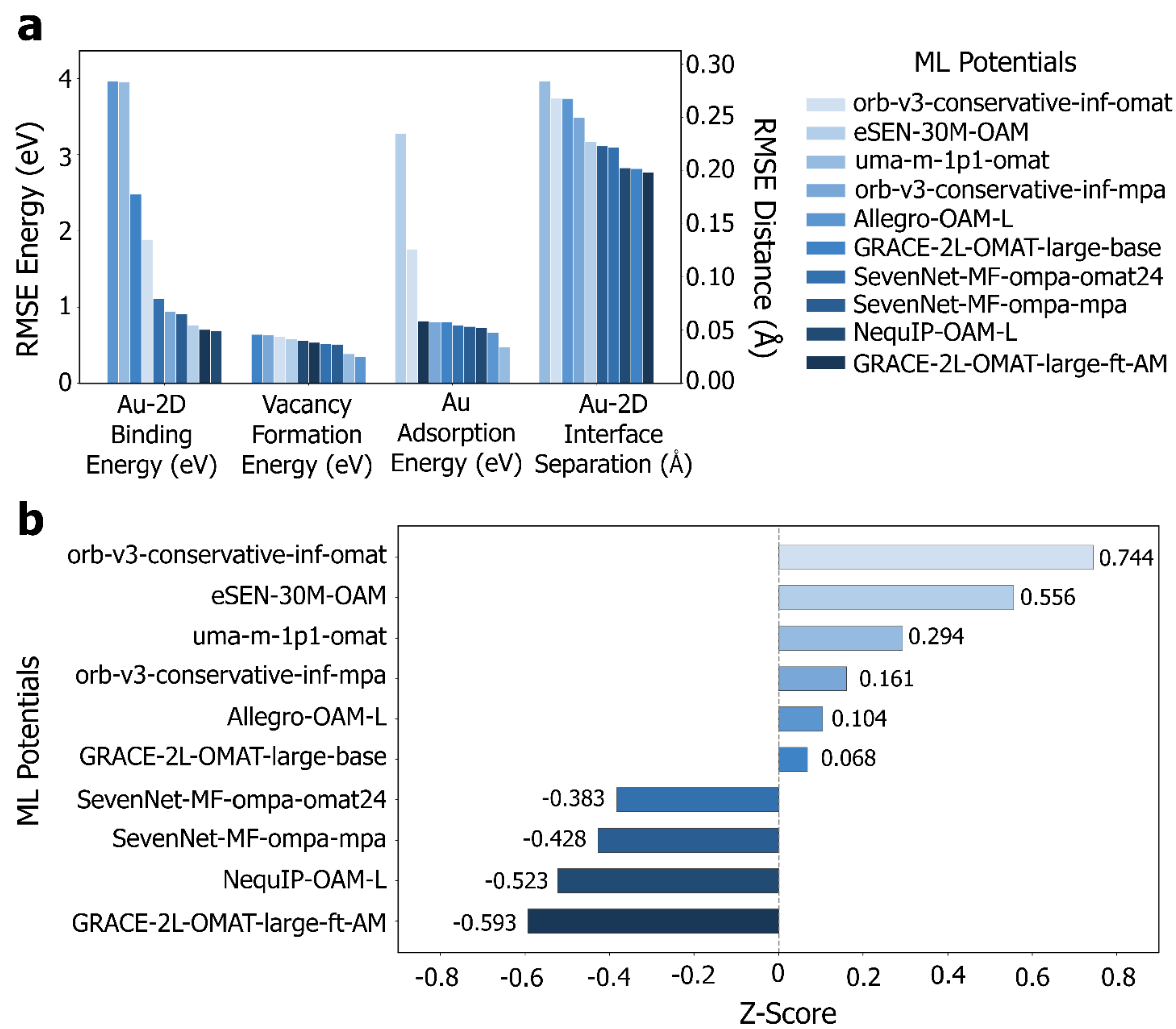

**Figure 7. Benchmarking of universal machine-learning interatomic potentials. (a)** Root-mean-square error (RMSE) for four key metrics, Au-2D binding energy, Au-2D interlayer distance, vacancy formation energy, and Au adsorption energy, evaluated for 17 candidate materials using ML potentials and benchmarked against DFT. **(b)** Z-score analysis was derived from the RMSE values for each potential, with more negative Z-scores indicating lower overall error and closer agreement with DFT across all metrics.

### 2.8 Kinetic spontaneity and the energetics of Au migration

To assess the physical feasibility of Au atom transfer between the electrode and vacancy site, we performed energy-barrier analyses using DFT-based nudged elastic band (NEB) calculations, complemented by MD simulations employing a universal machine-learning force field (MLFF). While such simulations do not directly yield the switching voltage, they enable a comparative evaluation of the likelihood of Au adsorption and desorption across different candidate materials. For a 2D material to function as an effective resistive memory, two stability conditions should be met. First, a significant forward migration barrier should exist to prevent the unsolicited adsorption of an Au atom into the vacancy, which would cause an unintended “SET” operation. Second, the reverse migration barrier should be high enough to inhibit instantaneous Au desorption from the vacancy back to the electrode, ensuring long-term retention of the programmed state.

Using NEB, we first compute the energy barriers for (i) adsorption of an Au atom from the Au electrode into the vacancy site of the 2D material (forward barrier) and (ii) desorption of the Au atom from the vacancy back to the Au electrode (reverse barrier). The NEB calculations are carried out by relaxing the initial and final images at 0K, followed by evaluation of the saddle-point energy along the minimum-energy pathway connecting these relaxed configurations. Barriers comparable to the thermal energy at 300 K (~0.026 eV) indicate a high likelihood of spontaneous Au migration. However, in some materials, the barrier height is marginally higher than 0.026 eV while remaining in a very low-energy regime (<0.1 eV). Thermal fluctuations at room temperature can alter the atomic positions, hence the energy landscape, such that migration may occur even for slightly higher barriers than 0.026eV. Consequently, it is essential to complement the NEB results with MD simulations at 300 K to directly evaluate the dynamical stability of the system. Performing DFT-based MD simulations for the large number of materials and configurations considered here is computationally prohibitive. Consequently, we employed a universal ML force field to enable long-time MD simulations. However, accurately capturing the energetics of metal-2D heterostructures requires a precise description of van der Waals (vdW) interactions, which are often underrepresented in standard ML potentials. Furthermore, since the universal ML potentials are typically not trained on point defects, surfaces, and 2D structures, it is crucial to perform a comparative analysis between energies obtained from the MLFF and those calculated using DFT, in order to validate the reliability of the ML-based MD simulations for the present system.

We initially selected the top-ranked ML interatomic potentials from the MatBench[44] Discovery platform, and we also tested out their close variants, such as MPA and OMAT24, trained versions of the same potentials (SevenNet and Orb). For GRACE, the two most accurate models were chosen. As the platform continuously updates its rankings in response to newly developed potentials and revised accuracy benchmarks, our selection is based on the rankings available in November 2025. The shortlisted potentials are summarized in **Figure 7**. To assess the accuracy of these ML potentials, we evaluated their performance using three energy metrics: Au-2D binding energies, vacancy formation energies, and Au adsorption energies, and one geometry metric: Au-2D equilibrium interface separations for all the shortlisted 17 materials and compared the results against DFT calculations. **Figure 7a** presents the root-mean-square error (RMSE) between energies and distances obtained from the ML potentials and DFT. For a more comprehensive comparison of model performance across multiple metrics, we performed a Z-score analysis (Methods), where a more negative Z-score indicates greater agreement with the DFT results (**Figure 7b**). Based on this analysis, the GRACE-2L-OMAT-large-ft-AM[49] and NequIP-OAM-L[50] models exhibit nearly identical Z-scores (**Figure 7b**). We therefore proceeded with GRACE-2L-OMAT-large-ft-AM, which shows a marginally more negative Z-score and thus the closest overall alignment with the DFT reference data, while being relatively fast computationally.

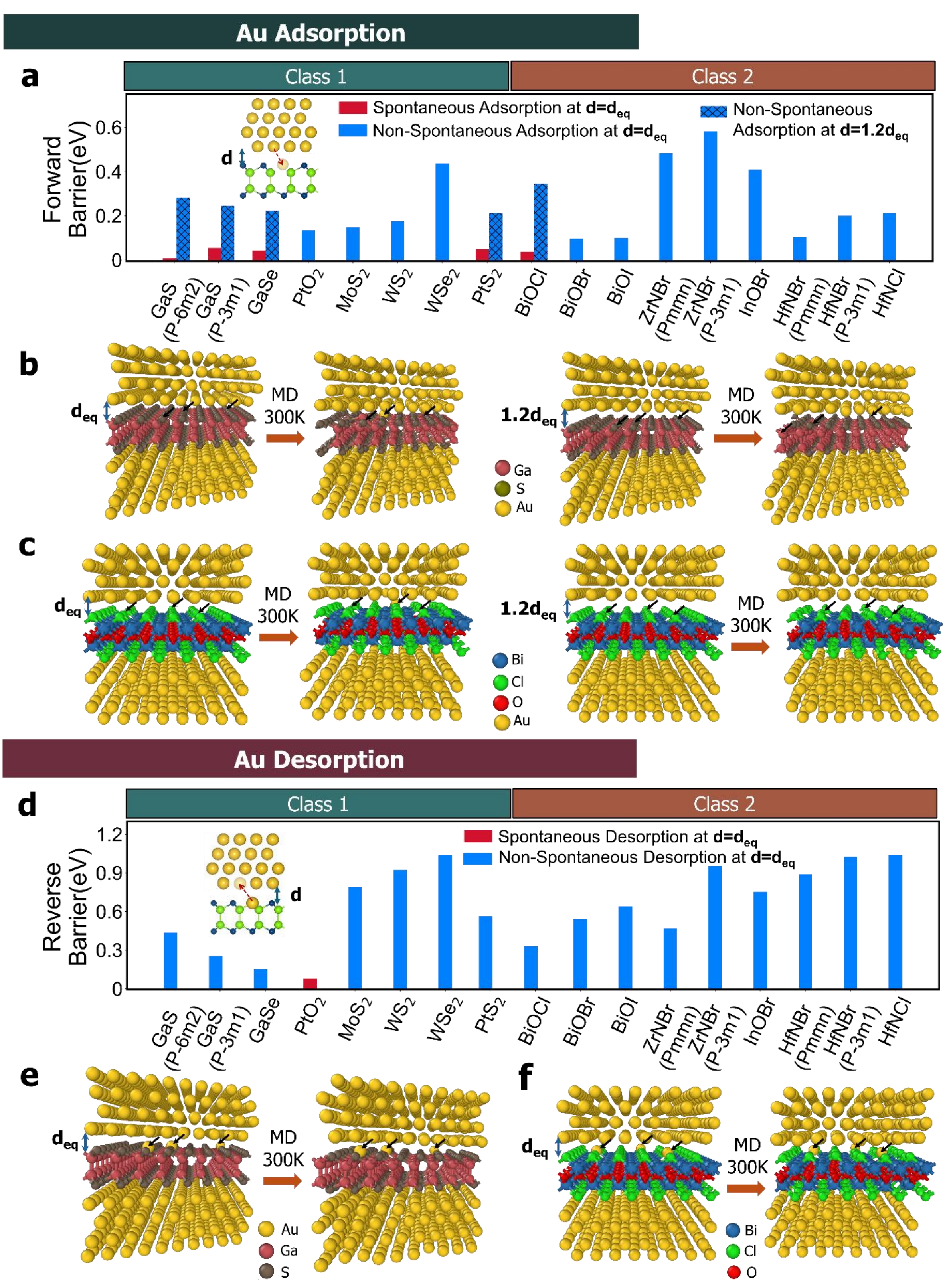

**Figure 8. Au adsorption and desorption feasibility study: (a)** DFT-NEB calculated energy barriers for Au atom adsorption from the electrode into vacancy sites (forward barriers) for all candidate materials. $d_{eq}$ denotes the equilibrium distance between Au and the 2D layer. The solid bars correspond to the equilibrium separation ($d_{eq}$), whereas the blue cross-hatched bars represent the barriers after increasing the metal-2D separation by 20% **(b)** Atomic snapshots at the onset (t = 0 ps) and end (t = 10 ps) of 300 K MD simulations of a representative Class 1 material, Au-GaS (P-6m2)-Au, at the equilibrium electrode-2D separation ($d_{eq}$) (left) and with the separation increased by 20% ($1.2d_{eq}$) (right). Three S vacancies are present in the 2D monolayer. At the equilibrium distance, the Au atom spontaneously migrates into an S vacancy without external stimuli, whereas increased separation suppresses spontaneous adsorption. **(c)** MD snapshots at 300 K for a representative Class 2 material, Au-BiOCl-Au, at the equilibrium separation (left) and with a 20%

increased separation (right). Similar to GaS, spontaneous Au occupation of the Cl vacancy occurs at the equilibrium distance but is inhibited when the separation is increased. **(d)** DFT-NEB calculated energy barriers for Au atom desorption from vacancy sites back to the electrode (reverse barriers) for all candidate materials. **(e)** MD snapshots at 300 K for Au-GaS (P-6m2)-Au with Au atoms initially placed in the vacancy sites and evolved for 10 ps at the equilibrium separation; the Au atoms remain trapped at the vacancies. **(f)** MD snapshots at 300 K for Au-BiOCl-Au with Au atoms initially placed in the vacancy sites. In contrast to adsorption, no spontaneous Au desorption into the electrode is observed during the simulation.

**Figure 8a** shows the NEB barriers for the adsorption of an Au atom from the electrode into the vacancy sites. At the equilibrium metal-2D separation ($d_{eq}$), obtained after DFT relaxation, the identified Class 1 materials (GaS-P6m2, GaS-P3m1, GaSe, and $PtS_2$) exhibit exceptionally low forward migration barriers. Although the barriers for GaS-P3m1, GaSe, and $PtS_2$ exceed the thermal energy at room temperature (~0.026 eV), MD simulations reveal that an Au atom spontaneously adsorbs at the vacancy sites in all four materials. Here, spontaneous adsorption is defined as an Au migration event occurring within the 10ps MD simulation window. In all observed cases, these events occur almost immediately, typically within the first 2ps of the trajectory before the total energy and temperature fluctuations reach a steady state, indicating the initial configuration is inherently unstable under ambient thermal conditions. Among the Class 2 materials, spontaneous adsorption is observed only for the Cl vacancy in BiOCl, whereas no adsorption events occur within the MD timescale for the remaining candidates. Accordingly, the bars in **Figure 8a** are color-coded based on the MD results, with red denoting spontaneous adsorption and blue indicating its absence. Notably, spontaneous adsorption is consistently observed in MD simulations when the NEB barrier falls below ~0.09 eV. While the static barrier height may shift under finite-temperature conditions, this threshold shows a clear correlation between the energy barrier and finite-temperature behaviour. We further observed that this adsorption behaviour is strongly dependent on the electrode-2D separation. Increasing the separation by more than 20% substantially increases the forward migration barriers (blue cross-hatched bars in **Figure 8a**), thereby suppressing spontaneous adsorption in all five materials.

Representative MD trajectories for GaS (P-6m2) and BiOCl at different separations are shown in **Figures 8b and 8c**. At $d_{eq}$, the Au atom readily occupies the S and Cl vacancies, whereas at larger separations it remains on the electrode throughout the 10 ps simulation. In **Figure 8d**, the reverse barrier for the desorption of an Au atom from the vacancy to the electrode is presented for all the candidates. Except for $PtO_2$, all candidates exhibit large reverse barriers, indicating strong kinetic stabilization of the Au-adsorbed state. MD simulations confirm that, for all materials except $PtO_2$, the Au atom remains stably trapped in the vacancy even at equilibrium distances. In contrast, for $PtO_2$, Au spontaneously returns to the electrode even at increased separations, indicating that the O vacancy cannot stably host the Au atom (**Supplementary Figure 11**). It should be noted that the MD simulations are primarily employed to detect rapid migration events that occur instantaneously upon interface contact. Consequently, these trajectories do not evaluate long-term state retention. A qualitative estimate of the characteristic residence time of vacancy and Au-adsorbed state can be estimated from the NEB barriers using the Arrhenius equation (**Supplementary Table 5**).

## 3. Discussions

Among the eight Class 1 monolayers identified, three ($MoS_2$, $WS_2$, $WSe_2$) have already demonstrated their potential in experimental atomristors[12,24], thus validating our computational screening workflow. The remaining five, comprising metal monochalcogenides and oxides, are previously unexplored, thus broadening the candidates for atomristor design. Although h-BN has been experimentally demonstrated in atomristors with HRS-to-LRS switching[20,22], it does not fall within our Class 1 monolayer candidate pool. In h-BN, an N vacancy introduces two defect states (one below the CBM and another above the VBM), yielding an effective gap of ~0.92 eV, which does not satisfy our band gap requirement of > 1 eV at the initial defective state. Furthermore, after Au adsorption, the conduction band edge moves toward the FL, but the effective gap remains ~0.68 eV (see **Supplementary Figure 2c**), which fails the < 0.5 eV threshold for LRS. However, the single-atomic-layer structure of h-BN enables a filamentary switching mechanism involving direct metal chain formation through vacancies.[51,52,26] This mechanism differs fundamentally from the vacancy-adsorption mechanism considered here and is expected to be suppressed in thicker monolayers (e.g., three-atom-thick $MoS_2$), where additional atomic sublayers hinder interelectrode metal migration. The distinct behavior of h-BN therefore points to an alternative class of ultrathin atomristors governed by direct metal-filament formation, which warrants further investigation.

The discovery of nine Class 2 monolayers (mainly metal oxyhalides and nitrohalides) unveils a previously unknown route to achieve an inverted atomristor capable of LRS-to-HRS switching. While similar LRS-starting behaviour has been reported in a bulk $TiO_2$ device[40], $TiO_2$ is more commonly employed in conventional HRS-to-LRS devices[53], and the underlying mechanism for such inverted switching has remained ambiguous. In standard bulk oxides or 2D multilayers, achieving stable inverted switching is inherently challenging, as resistive switching is typically governed by metal filament formation that short-circuits the electrodes and enhances conductivity[29]. The atomically thin nature of monolayer atomristors enables a fundamentally different, defect-mediated switching mechanism, making the LRS-to-HRS transition a significant advance in atomristor design. Because Class 2 materials are initially in an LRS, they eliminate the need for a high-voltage initial “SET” pulse, making them ideal for RESET-dominated workloads[54] in artificial neural networks (ANNs). The functional complementarity[41,55] of Class 1 and Class 2 atomristors can potentially be integrated into crossbar arrays, where one class can conduct while the other suppresses current, thus mitigating sneak-path currents without additional selector elements. Furthermore, this pairing naturally may provide a route to mimic the excitatory-inhibitory balance[43] of biological systems, where Class 1 elements drive potentiation and Class 2 elements drive depression. Moreover, several Class 2 materials identified here, such as BiOCl, ZrNBr, and ZrNCl, have also been recognized as excellent dielectric materials, exhibiting low leakage currents in field-effect transistors.[56,57] Since excessive leakage can undermine switching reliability, integrating such monolayers could lead to energy-efficient and high-fidelity memory architectures.

Beyond identifying new atomristor materials, our results provide a device-engineering strategy based on controlling the electrode-2D interfacial distance. Quantum transport simulations reveal that the ON/OFF ratio is strongly dependent on the electrode-2D interfacial distance. At the equilibrium separation, the ON/OFF ratio is too small to support reliable resistive switching (**Supplementary Figure 6**), consistent with prior studies attributing this limitation to MIGS[13]. Recent simulations on $MoS_2$ and h-BN atomristors show that the ON/OFF characteristics are extremely sensitive to electrode geometry and metal-2D interlayer distance[13,32,58]. This dependence may also explain the wide variation in experimentally reported ON/OFF ratios for identical materials[12,13,15], as fabrication-induced residues can modify the effective electrode-2D separation[58]. For the monolayers explored in this work, i.e. GaS(P-6m2), GaS(P-3m1), GaSe, $PtS_2$ and BiOCl, spontaneous Au adsorption occurs at the equilibrium electrode-2D separations. Increasing the separation suppresses this spontaneous adsorption and therefore provides a practical route toward voltage-triggered switching. However, this improvement in transport and switching stability comes with a kinetic trade-off: larger electrode-2D distances increase the Au migration barrier (**Supplementary Figure 9**) and may consequently require higher switching voltages**.** The newly identified Class 1 materials are particularly attractive because they possess exceptionally low migration barriers at equilibrium. Hence, the interfacial distance can be increased to enhance the resistance window without incurring prohibitively large switching barriers, thereby maintaining relatively low switching voltages and enabling a high ON/OFF resistance ratio. Furthermore, their reverse migration barriers remain comparable to those of experimentally established ($MoS_2$, $WS_2$, and $WSe_2$) materials (**Supplementary Figure 10**), thus indicating that RESET operation remains feasible at increased separations. These results demonstrate that electrode-2D separation can serve as a device-level tuning parameter to optimize the resistance window, switching voltage, and stability of the resistive states. The contrasting behavior of $PtO_2$, where spontaneous Au desorption persists even at increased separation (**Supplementary Figure 11**), further highlights the importance of combining materials selection with interface engineering. While unsuitable for non-volatile memory, such behaviour may be advantageous for volatile or threshold-switching applications.[26]

NEB barrier analysis reveals that several Class 2 materials, such as BiOBr, BiOI, HfBrN (Pmmn), HfBrN (P-3m1), and HfNCl, exhibit low forward migration barriers together with relatively high reverse barriers (**Figure 8**). This combination is desirable for RS devices, as it implies lower forward switching voltage and enhanced retention of the Au-adsorbed state.[35] In **Supplementary Table 5**, we estimate characteristic residence times of both the vacancy state and the Au-adsorbed state using transition-state theory based on the NEB barriers at room temperature across a standard trial frequency range ($10^{12}$-$10^{13}$ $s^{-1}$). Among the experimentally demonstrated benchmark atomristor materials, $WS_2$ exhibits the largest reverse barrier (1.04 eV), followed by $MoS_2$ (0.92 eV), corresponding to the longest characteristic residence times of the Au-adsorbed state. Among the identified candidates, the Class 1 materials GaS (P-6m2), GaS (P-3m1), as well as the Class 2 materials ZrNBr (P-3m1), HfNBr (P-3m1), and HfNCl exhibit reverse barriers comparable to the benchmark materials. In particular, HfNCl and HfNBr (P-3m1) exhibit reverse barriers above 1 eV, corresponding to an isolated Au-adsorbed state residence time of $> 2 \times 10^4$s. It should be emphasized that these residence times are idealized Arrhenius estimates based on a single local migration path. Consequently, they represent the characteristic timescale for an isolated Au migration event in a simplified electrode-2D configuration and should be interpreted as comparative indicators of migration kinetics among different materials rather than quantitative predictions of macroscopic device-level retention. In an operational memristor, experimental retention is a collective property additionally influenced by finite-temperature fluctuations, multi-defect interactions, and the surrounding electrode architecture.

## 4. Conclusions

In summary, we presented a comprehensive computational framework to discover, classify, and rationally design 2D monolayers for RS devices. Our findings expanded the Class 1 landscape beyond metal dichalcogenides to include several metal monochalcogenides with superior switching kinetics. Class 2 monolayers offered a previously unexplored inverse switching mechanism at the atomic scale, thus opening new opportunities for low-power RESET-dominated and inhibitory device functionalities. The identification of two distinct functional classes diversifies the switching operation of atomristor from single-mode into a complementary dual-mode fashion, thus establishing a material basis for streamlining the design of unconventional computing architectures based on 2D materials and atomristors. Our proposed framework can be extended to different metal electrodes and few-layer systems, thus providing a computational design basis for the rapid prediction of switching kinetics and electronic modulation across diverse 2D-metal interfaces and devices.

## 5. Methods

### 5.1 Defect Formation

For defect formation, $4\times4\times1$ supercells are built for each material, and vacancy defects are generated in the supercell using the *doped*[59] automated Python package. To simplify the simulations, we focused exclusively on neutral vacancies. Charged defects were not included in this study, as our primary objective was to assess the changes in the density of states (DOS) induced by Au adsorption. Neutral defects are sufficient to capture the relevant electronic structure modifications without introducing the added complexity associated with charged states.

The vacancy formation energies were calculated using **Equation 1**.

$$\Delta E_f^V = E_{tot}^{Vacancy} - E_{tot}^{Pristine} + \mu_X \quad (1)$$

where, $\Delta E_f^V$ represents the vacancy formation energy. $E_{tot}^{Pristine}$ and $E_{tot}^{Vacancy}$ denote the total energy of the pristine and vacancy-embedded supercell. $\mu_X$ signifies the chemical potential of the element ($X$) removed to form the vacancy. It is important to note that our vacancy formation energy calculations were performed without assuming any specific chemical environment (e.g., metal-rich, oxygen-rich, or halogen-rich conditions). Including such factors would have introduced immense complexity, given the significant number of materials in our study. Instead, we established a standardized reference point: the chemical potential for each vacancy constituent element was determined based on the total energies of their most stable elemental bulk phases. **Supplementary Table 3** lists the corresponding bulk structures and chemical potential for each element. This approach provides a consistent and material-agnostic baseline for comparing vacancy formation tendencies across a diverse set of compounds, without biasing the results toward any particular synthesis condition. This energy-based method offers a thermodynamic perspective on defect stability under equilibrium conditions.

### 5.2 DFT Calculations

DFT calculations are carried out using Vienna Ab-initio Simulation Package (VASP) with the projected augmented-wave (PAW) method[60,61]. Since atomic defects can induce or modify spin properties, spin-polarized calculations are essential for accurately determining the electronic characteristics of these systems. However, widely used DFT approaches, such as the local density approximation (LDA) and generalized gradient approximation (GGA), suffer from significant self-interaction errors (SIE)[62], particularly in magnetic systems, leading to inaccurate predictions of electronic band structures. A common approach to address this issue is by applying on-site empirical corrections to localized orbitals, such as the Hubbard U correction (GGA + U). However, the U parameter is highly material-specific, making it impractical to determine for every defective configuration in a high-throughput setting.[63] Hybrid functionals provide greater accuracy, but their computational cost is prohibitive for large-scale studies involving extensive supercell calculations. Meta-GGA functionals, such as r2SCAN (Revised Regularized Strongly Constrained and Appropriately Normed)[47], represent a promising alternative[64]. They enhance d-electron localization and yield more reliable band gap predictions, offering a substantial improvement over conventional LDA and GGA functionals. Recent research indicates that the r2SCAN functional has proven to be an efficient and accurate method for calculating various properties in magnetic materials[65,66]. Therefore, we adopted the r2SCAN functional for all DFT calculations in this study, including the structural relaxation of the defective supercells. For the relaxation of the material unit cell, we used $\frac{30}{a} \times \frac{30}{b} \times 1$, while for DOS calculations we used $\frac{60}{a} \times \frac{60}{b} \times 1$ gamma-centered K-point grid. Where $a$ and $b$ are lattice parameters of the 2D materials. The unit cells are relaxed unless maximum Hellmann-Feynman forces on each atom are less than 0.01 eV/Å. The BZ is integrated with Gaussian smearing using a well-tested smearing factor of 0.05 eV. A large vacuum space of about 20 Å is placed in the direction of *c* to avoid spurious interaction between periodically repeated layers. Dipole corrections are introduced in the z direction.

Because a large number of spin-polarized calculations on defective supercells would be computationally expensive, we loosen the relaxation and SCF criteria compared to the unit cell simulation. The defective supercells are relaxed with a gamma k-point, and energy and force convergence criteria are taken as 1e-4 and 0.04 eV/Å. For density of states calculations, $\frac{60}{a} \times \frac{60}{b} \times 1$ K-points are chosen, and the energy convergence criteria are set at 1e-5. The Bader charge analyses are performed using the code developed by Henkelman et al.[67], where charge densities generated from DFT static runs are used as inputs. For the interface calculations of Au-GaS, 5×5×1 supercell <111> plane of Au is stacked with a 4×4×1 supercell of GaS using Atomic Simulation Environment (ASE)[68]. We use the *py4vasp* Python interface to extract data from VASP calculations.

### 5.3 Molecular Dynamics Simulations

#### 5.3.1 Defect Stability Assessment

Molecular dynamics simulations were conducted using the ASE package[68], and visualizations of the simulation results were performed with OVITO[69] and VESTA[70]. For the initial defect stability study, the interatomic forces were described using a universal machine learning potential Orb-v3[48] potential trained on the OMAT database (model orb_v3_conservative_inf_omat). Orb-v3 is chosen for the first phase of MD simulations as it was within the top 10 potentials in MatBench[44] Discovery and computational efficiency are better. For the MD simulations, Newton's equations of motion were integrated with a time step of 1 fs. The MD simulations were carried out at two separate temperatures, 300 K and 700 K, for 20ps. We used Berendesen NPT dynamics to maintain constant ambient pressure and temperature throughout the simulations. For these MD simulations, we constructed a 15×15×1 supercell of the chosen 2D materials. To ensure the stability of the simulated materials, even under conditions that might arise from inefficient fabrication, we introduced a relatively high concentration of vacancies. An automated Python script was used to create the most probable vacancies throughout the supercell with a 10% vacancy concentration.

#### 5.3.2 Benchmarking ML Potentials

To ensure the accuracy of the Au adsorption-desorption kinetics, we evaluated the performance of ten universal ML potentials against DFT data. To maintain consistency with the ML training sets, all benchmark DFT calculations utilized the GGA-PBE functional, with vdW interactions accounted for via the Becke-Johnson (BJ) damping scheme in both frameworks. The following four physical metrics were used for the comparative RMSE (Root Mean Square Error) analysis:

1. Interface Binding Energy ($E_B$): First, Au-2D heterostructures were constructed by placing a four-layer Au slab on top of the 2D monolayer using the Interface Builder module in QuantumATK. The supercells were formed to minimize the lattice strain below 1%. Depending on the material, either Au(111) or Au(100) surfaces were selected to achieve optimal lattice matching and minimize interfacial strain. The structures were subsequently relaxed using both DFT and the ML potentials. Then, the binding energy was calculated using Equation 2:

$$E_B = E_{Au-2D} - (E_{Au} + E_{2D}) \quad (2)$$

where, $E_{Au-2D}$ , $E_{Au}$ and $E_{2D}$ represent the total energies of the combined interface, the isolated Au electrode, and the 2D monolayer, respectively.

2. Interfacial Separation ($d$): Determined via a custom script that calculates the distance between the minimum z-coordinate of the Au electrode and the maximum z-coordinate of the 2D material.
3. Vacancy formation energy ($\Delta E_f^V$): Evaluated according to Equation 1.
4. Au adsorption energy ($\Delta E_{ads}^{Au}$): Calculated using Equation 3.

$$\Delta E_{ads}^{Au} = E_{tot}^{Vac+Au} - E_{tot}^{Vac} - E_{Au} \quad (3)$$

where $E_{tot}^{Vac_Au}$ is the total energy of the system with an Au atom occupying the vacancy site, $E_{tot}^{Vacancy}$ is the total energy of the defective 2D monolayer, and $E[Au]$ is the energy of an isolated Au atom. These vacancy-related calculations were performed without constructing an explicit Au-2D interface. 4×4×1 supercell was used for all vacancy formation and Au adsorption calculations, containing a single vacancy and a single Au atom.

To standardize the performance metrics across the 10 evaluated machine learning potentials, we calculated the Z-score for each RMSE value.

$$Z_{i,j} = \frac{RMSE_{i,j} - \mu_j}{\sigma_j} \quad (4)$$

where $i$ and $j$ define indices of ML potentials ($i$ = 1,2,3…10) and four metrics ($j$ = 1,2,3,4). $\mu_j$ and $\sigma_j$ are the mean and standard deviation of RMSE values across all potentials for each metric $j$, where

$$\mu_j = \frac{1}{10}\sum_{i=1}^{10} RMSE_{i,j},\ \sigma_j = \sqrt{\frac{1}{10}\sum_{i=1}^{10}\left(RMSE_{i,j} - \mu_j\right)^2}. \quad (5)$$

**5.4 Quantum transport simulations**

The current under different bias are computed using the non-equilibrium Green's function (NEGF) method and DFT as implemented in the QuantumATK code[71]. The geometries used in the DFT-NEGF calculation are shown in the Supplementary Figure 8. DFT calculations are done using a linear combination of atomic orbitals (LCAO) approach, utilizing a Double-Zeta Polarized (DZP) basis set and an 80 Ha energy cutoff. A Monkhorst-Pack k-point grid of 3 × 3 × 150 is used. For vdW interaction, Grimme DFT-D3 van der Waals correction is applied. For finite bias calculation, positive or negative bias $+V$ or $-V$ has been applied to the top electrode while the bottom electrode is kept at zero bias. The current under a bias $V$ is derived from the Landauer formula as given below[71].

$$I(V) = \frac{2e}{h}\int_{-\infty}^{+\infty} T(E,V)[n_F(E-\mu_L) - n_F(E-\mu_R)]\, dE$$

where, $h$ is Planck's constant, $T(E,V)$ is the transmission function at energy $E$ and voltage $V$, $n_F$ is the Fermi distribution function, $\mu_L$ and $\mu_R$ are the chemical potentials of the two electrodes.

## Acknowledgments

This work was supported by the Singapore National Research Foundation (NRF) Frontier Competitive Research Programme (F-CRP) under the award number NRF-F-CRP-2024-0001. H.Z. is supported by the National Natural Science Foundation of China (Nos. 12004307). The computational work for this article was fully performed on resources of the National Supercomputing Centre, Singapore (https://www.nscc.sg).